# The Electrostatic Screening Length in Concentrated Electrolytes Increases with Concentration

Alexander M. Smith*,[a], Alpha A. Lee*,[b] and Susan Perkin*,[a]

[a]Department of Chemistry, Physical & Theoretical Chemistry Laboratory, University of Oxford, Oxford OX1 3QZ, U.K.
[b]School of Engineering and Applied Sciences, Harvard University, Cambridge, MA 02138, USA

Corresponding author emails:

susan.perkin@chem.ox.ac.uk
alphalee@g.harvard.edu
alexander.smith@chem.ox.ac.uk

**ABSTRACT**

According to classical electrolyte theories interactions in dilute (low ion density) electrolytes decay exponentially with distance, with the *Debye screening length* the characteristic length-scale. This decay length decreases monotonically with increasing ion concentration, due to effective screening of charges over short distances. Thus within the Debye model no long-range forces are expected in concentrated electrolytes. Here we reveal, using experimental detection of the interaction between two planar charged surfaces across a wide range of electrolytes, that beyond the dilute (Debye-Hückel) regime the screening length *increases* with increasing concentration. The screening lengths for all electrolytes studied – including aqueous NaCl solutions, ionic liquids diluted with propylene carbonate, and pure ionic liquids – collapse onto a single curve when scaled by the dielectric constant. This non-monotonic variation of the screening length with concentration, and its generality across ionic liquids and aqueous salt solutions, demonstrates an important characteristic of concentrated electrolytes of substantial relevance from biology to energy storage.

**TOC Image:**

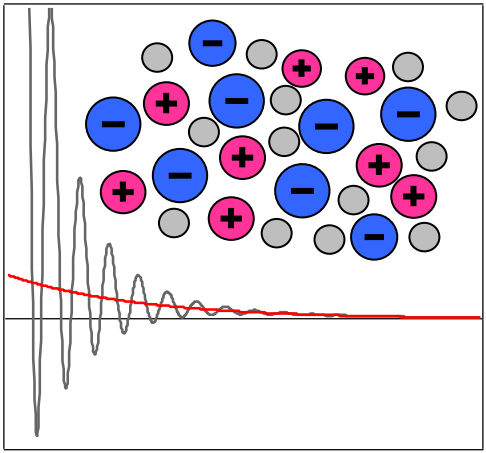

Electrolytes are ubiquitous in nature and in technology: from the interior of cells to the oceans, from supercapacitors to nanoparticle dispersions, electrolytes act as both solvent and ion conduction medium. The 'high concentration' end of the electrolyte spectrum is of particular relevance; examples of natural and technological electrolytes include human blood plasma (~ 0.15 M), the Atlantic Ocean (~ 0.6 M), supercapacitor electrolytes (1-2 M), and the Dead Sea (~ 4.7 M). Despite this, the properties of concentrated electrolytes remain relatively uncharted territory, particularly from an experimental point of view, compared to the well-understood dilute electrolyte regime where Poisson-Boltzmann (mean-field) electrostatics applies and approximations such as in the Debye-Hückel theory are appropriate[1-4].

Within the classical mean-field theories that explain screening and interactions in dilute electrolytes[1], the Debye length, $\lambda_D$, is the yardstick against which the reach of electrostatic interactions in solution is typically measured. The Debye length scales inversely with the square root of bulk ion density, $\rho_\infty$, according to:

$$\lambda_D = \left(\frac{\varepsilon_r \varepsilon_0 k_B T}{\sum_i \rho_{\infty i} e^2 z_i^2}\right)^{1/2}$$

where $\rho_{\infty i}$ is the number density of ion type $i$, $\varepsilon_r$ is dielectric constant, $\varepsilon_0$ is the permittivity of free space, $T$ the temperature, $k_B$ the Boltzmann constant and $z_i$ is the ion valency. Thus, within mean-field Poisson-Boltzmann models, a high concentration (or density) of ions leads to screening of an electrostatic potential over short distances, whilst low ion densities allow the electric potential to decay over longer distances and so the electric field (gradient of potential) to act over greater distances from a charged particle or object. For example a solution of NaCl in water at 25 °C and a concentration of $1 \times 10^{-3}$ M has a Debye length of $\lambda_D$ = 9.6 nm, corresponding to roughly 40 water diameters; whereas at a concentration of 0.1 M the Debye length is $\lambda_D$ = 0.96 nm, equivalent to just four water diameters. At higher concentrations, when the Debye length becomes comparable to molecular dimensions, the Debye-Hückel theory is not expected to apply[5]; ion size (steric or volume exclusion) effects and ion correlations come into play. Models to account for this include extensions based on the Poisson-Boltzmann equation[6-7], or analysis of the Ornstein-Zernike equation for binary ionic fluids[3, 8], leading to the prediction of oscillatory (rather than monotonic) decay for concentrated electrolytes[9]. Although it is well known within the theoretical community that the asymptotic decay length is determined by either charge-charge or hard core correlations and could exceed the Debye length in concentrated electrolytes[8, 10], this has not been directly measured in electrolytes in the past and its implications are not widely appreciated.

We report here measurements of the interaction force between two smooth planar surfaces across electrolytes showing an exponentially decaying component to the interaction force extending over long range even at high electrolyte concentrations, i.e. the experimental decay length greatly exceeds the theoretical Debye length; $\lambda_{exp} \gg \lambda_D$. This observation is in accordance with a recent observation of long-range screening in ionic liquids[11]; we now provide measurements of a wide range of pure ionic liquids, ionic liquids dissolved in a polar solvent (propylene carbonate), and simple NaCl salt in water to demonstrate this to be a general observation of electrolytes at high concentration.

We uncover a clear and strong non-monotonic relationship between the exponential decay length of the observed long range force, $\lambda_{exp}$, and the ion density (concentration), as follows. In dilute electrolytes the decay length decreases with concentration as expected within the Debye-Hückel regime, reaching unmeasurably small $\lambda_{exp}$ at ~ 0.1M. Then, as the electrolyte concentration is increased further, beyond ~ 0.5M, we find that $\lambda_{exp}$ *increases* again with increasing concentration, up to maximum values greater than the theoretical Debye length by a factor of ~120. This trend is followed by all electrolytes studied including ionic liquids diluted with a polar solvent (propylene carbonate) and NaCl solutions in water. Scaling the measured decay length of interactions in each electrolyte or ionic liquid by its respective dielectric constant allows us to collapse of the data onto a single curve, suggesting that these observations are independent of molecular architecture or details of molecular interactions and instead are a general feature of concentrated electrolytes with a common underlying mechanism.

Measurements of interaction force between two atomically smooth and large-area mica sheets in crossed-cylinder configuration, across the electrolytes of interest, were carried out using a Surface Force Balance (SFB)[12]. The apparatus and procedure is essentially similar to that used in early measurements of DLVO forces[13] and experiments with pure ionic liquids[11, 14]. In brief: the back-silvered mica sheets create an interferometric cavity so that the absolute separation distance between them can be measured using white light interferometry (measuring the Fringes of Equal Chromatic Order, FECO). The electrolyte is held in a droplet between the crossed-cylindrical lenses, such that the confined film is in equilibrium with a large excess of bulk liquid outside of the confined film. Interaction forces arising from the difference in pressure in the film and the bulk lead to the deflection of a spring upon which one of the mica sheets is mounted, and this is detected in the spectral fringes. All measurements were carried out at T = 294K. Electrolytes (details in the supporting information) were injected between the two freshly cleaved mica sheets and forces, $F_N$, were measured as a function of film thickness (or mica surface separation), *D*. In our previous works with ionic liquids we focussed on the short-range oscillatory forces; here we show results of high resolution measurements at long range allowing us to observe both the oscillatory region and an exponentially decaying force that exists at distances beyond the oscillatory part of the interaction. Further details of the experimental procedure are in the supporting information.

We begin by presenting and discussing the nature of the mica-mica interaction across concentrated electrolyte for two example cases: one pure ionic liquid (intrinsic concentration of the pure liquid = 3.3 M) and one solution of NaCl in water at 2.0 M. Figure 1A shows, for an example ionic liquid 1-butyl-1-methylpyrrolidinium *bis*[(trifluoromethane)sulfonyl]imide, [$C_4C_1$Pyrr][NTf$_2$], the total force law between negatively charged mica surfaces across the liquid as a function of the liquid film thickness. The data is plotted on a log-linear scale in the main panel, and the short-range part of the same data on a linear-linear scale in the inset. An oscillatory component to the force is apparent in the region 0 – 6 nm; this has been the focus of our previous studies[15] and we do not discuss it further here. Beyond the oscillatory region, from ~ 6 – 20 nm, we measure a monotonic and exponentially decaying repulsive force. The decay length of the exponential region is the 'experimental decay length', $\lambda_{exp}$, of the long-range force, in this case ~8.4 nm, as discussed and compared with equivalent values for a range of electrolytes in the following discussion.

Figure 1B shows a similar force law measured for a concentrated aqueous electrolyte: 2M NaCl in water. Again, an oscillatory region is measured at short range, in this case with an oscillatory period of ~0.5nm and extending up to distances of 3 nm, followed by an exponentially decaying monotonic region from ~ 3 – 6 nm with $\lambda_{exp}$ = ~1.1 nm. The forces within the oscillatory region, 0 – 3nm, have been discussed in the past and are thought to be caused by squeeze-out of sequential layers of hydrated (or partially hydrated) ions[16] and will not be interpreted further here. We note, however, that detection of these oscillatory forces in NaCl are a crucial signature of cleanliness of the system; experiments where these were not observed were often found to be irreproducible between different contact areas and were discarded due to contamination. The monotonic part, in the region 3 – 6 nm, has also been noted in studies of forces across other concentrated aqueous electrolytes in the past[17-19].

We have conducted measurements similar to those in Figure 1 for a range of electrolytes at different concentrations, all measured at T = 294 K. In each case the asymptotic decay (manner in which the force decays towards zero at large *D*) was well represented by an exponentially decaying function, i.e. for separations sufficiently far beyond the oscillatory regime $F_N/R \sim \exp(-D/\lambda_{exp})$, with $\lambda_{exp}$ a characteristic 'experimental decay length' different for each electrolyte. In figure 1 the values of

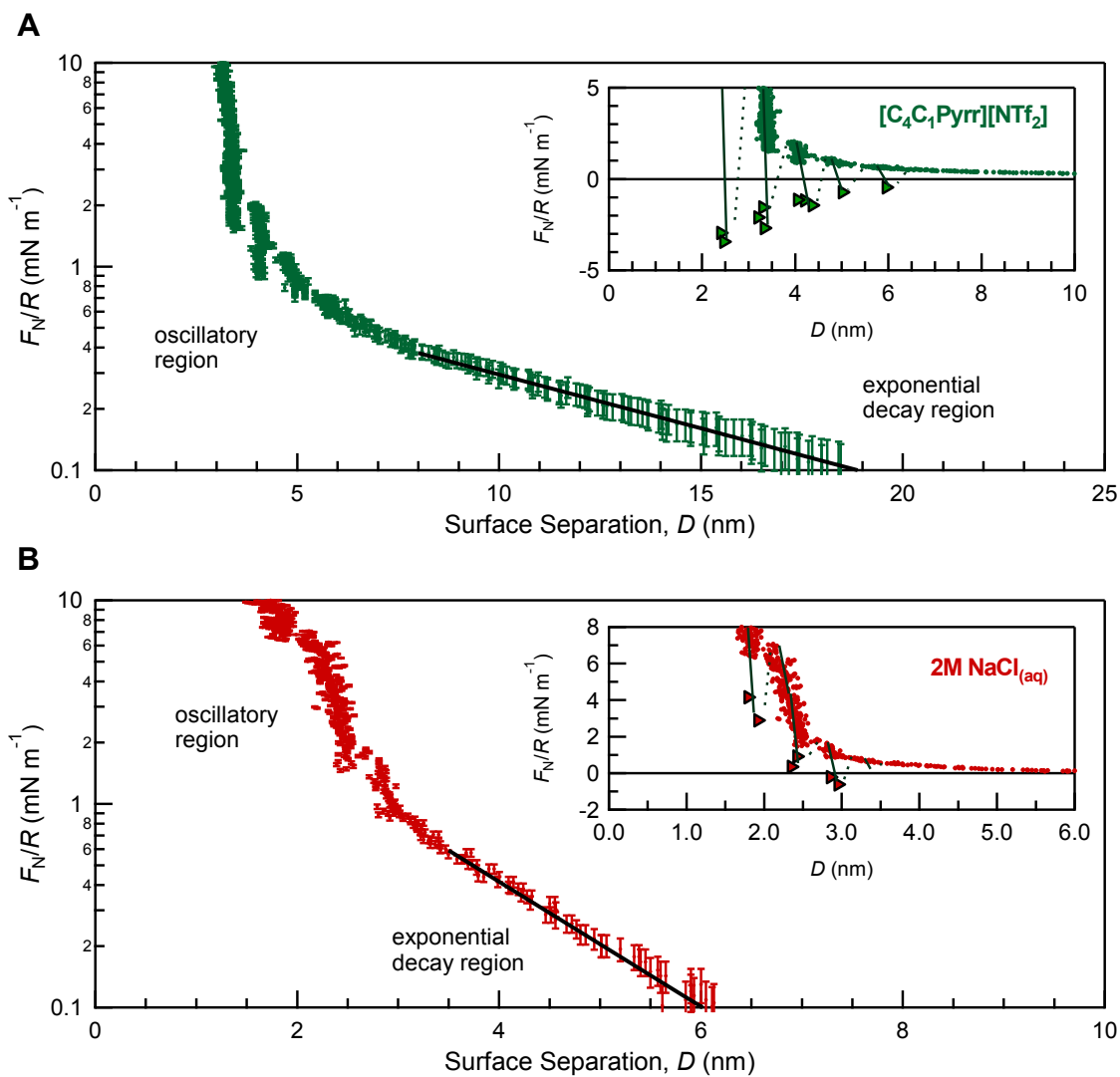

**Figure 1**
The measured normal force, $F_N$, normalised by radius of curvature, $R$, between crossed cylindrical mica surfaces across (A) pure [$C_4C_1$Pyrr][$NTf_2$] ionic liquid and (B) 2M $NaCl_{(aq)}$. Shown on log scale are the forces measured on approach of the surfaces, with exponential fits of the long range force shown as solid lines. In each case the insets show, on linear scales, the combination of forces measured on approach (dots) and retraction of the surfaces (triangles) which reveals the nature of the oscillatory region at small distances.

$\lambda_{exp}$ are the gradients of the black lines. We note here that the asymptotic decay length of the force between charged surfaces across a fluid is known to match the bulk asympototic decay in the same fluid[8], and so in this experiment we are probing in a systematic and high-resolution manner a property of bulk electrolyte. This decay length, $\lambda_{exp}$, can therefore be thought of as an experimentally determined screening length for interactions in the electrolyte. In the following sections we compare and discuss $\lambda_{exp}$ for simple NaCl in water and an ionic liquid in propylene carbonate – systematically varying the concentration in each case – and for a range of different pure ionic liquids.

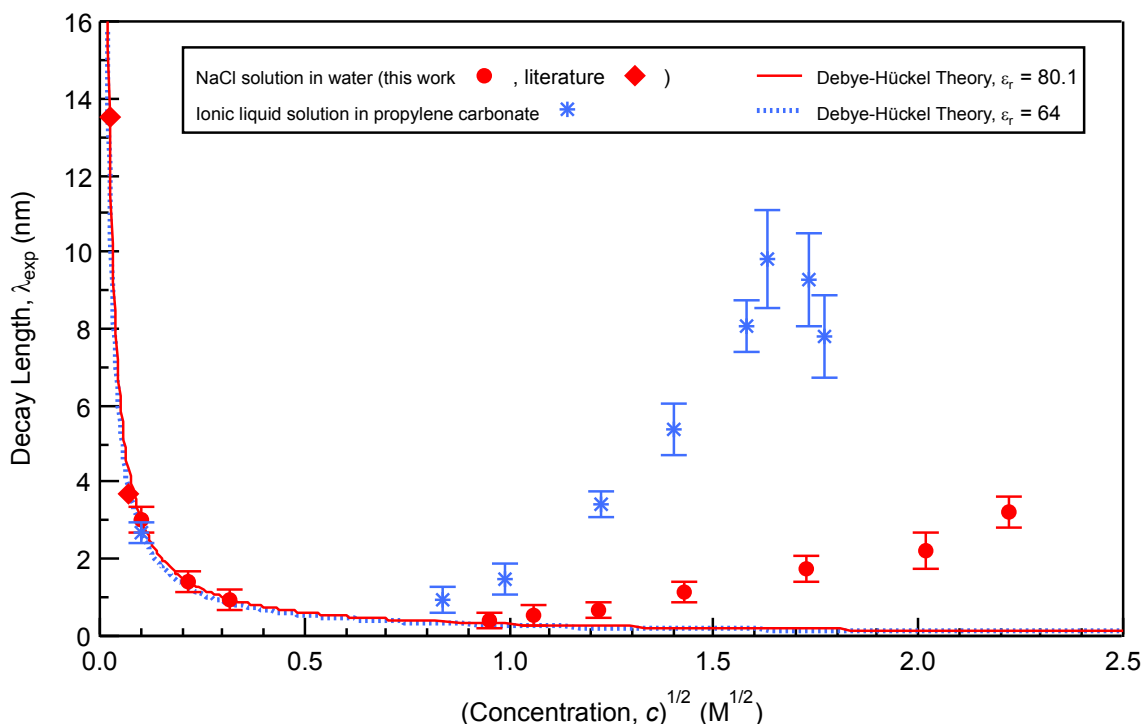

**Figure 2**
The measured decay length, $\lambda_{exp}$, plotted as a function of $c^{1/2}$, for aqueous NaCl solutions (filled circles), and solutions of an ionic liquid (IL), [$C_4C_1$Pyrr][$NTf_2$], mixed with propylene carbonate (asterisk symbols) and compared to values in the literature[13] (filled diamonds). The solid and dashed lines show how the theoretical Debye length varies with $c^{1/2}$ for two values of dielectric constant, $\varepsilon_r$ = 80.1 (solid line) and $\varepsilon_r$ = 64 (dashed line). Details of the electrolytes, including the ionic liquid chemical structure, are in the supporting information.

Figure 2 shows $\lambda_{exp}$ plotted as a function of $c^{1/2}$, for ease of comparison to the theoretical Debye length $\lambda_D$, for ionic liquid diluted with propylene carbonate and NaCl salt dissolved in water. Also marked on Figure 2 are lines indicating the expected (theoretical) decay length, $\lambda_D$, for two values of $\varepsilon_r$; $\varepsilon_r$ = 80.1 (appropriate for NaCl in water at low concentration) and $\varepsilon_r$ = 64.0 (pure propylene carbonate). We note that here $\lambda_D$ is calculated according to Equation 1 with the assumption that all ions are free, i.e. $\rho_\infty$ is the total concentration of ionic species; we do not propose all ions *are* indeed 'free', but simply that this is a suitable reference length-scale with which to measure deviation from ideal (dilute electrolyte) behaviour.

It is clear by inspection of Figure 2 that $\lambda_{exp} \approx \lambda_D$ at low concentrations ($c$ < 0.1 M; although this depends on dielectric constant as we show later), as has been demonstrated many times in the past[1], whereas at high concentrations there is strong deviation of $\lambda_{exp}$ from $\lambda_D$. We make the following observations: (i) There appears to be a concentration of minimum $\lambda_{exp}$ in the range $c_{min}$ ~ 0.1 – 0.5 M; and (ii) above this concentration $\lambda_{exp}$ *increases* with concentration, showing strong quantitative and qualitative deviation from the theoretical Debye length. The magnitude of the measured $\lambda_{exp}$ is substantially lower for NaCl in water than for the propylene-carbonate electrolyte; this is accounted for by their different dielectric constants as discussed later.

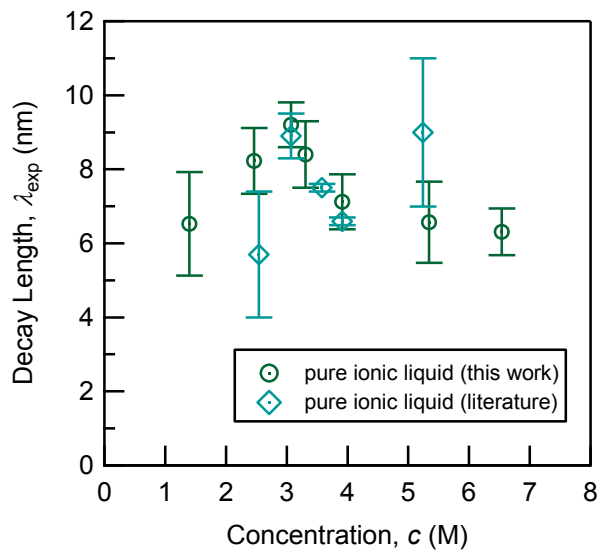

**Figure 3**
The measured decay length, $\lambda_{exp}$, plotted as a function of molar ionic liquid concentration, $c$, for different pure ionic liquids studied here (circles) and compared to values in the literature[11, 20-21] (diamonds). The chemical details of each ionic liquid, and the individual decay and concentration data corresponding to each, are given in the supporting information.

We also studied pure ionic liquids and the resulting $\lambda_{exp}$ values are shown in Figure 3. Variation in the intrinsic concentration for pure ionic liquids arises from differences in molecular volume and packing; details of the individual ionic liquids studied and their intrinsic concentrations are in the supporting information. In every case a long-range monotonic repulsion was observed with decay-length substantially greater than the theoretical value, i.e. $\lambda_{exp} \gg \lambda_D$ (where $\lambda_D \sim 0.1$ nm for these pure ionic liquids). Also plotted in Figure 3 are data points taken from other authors[11, 20-21] showing good agreement with respect to magnitudes of $\lambda_{exp}$. The scatter in values for pure ionic liquids in Figure 3 is at least partially attributed to their varying ion sizes and dielectric properties, as described later and shown in Figure 4.

When considering the origin of the longest range decay we arrive first at the question of the extent to which molecular (chemical) features of the ions play a role, or whether the behaviour can be explained purely in terms of the electrolyte concentration and dielectric constant. To illuminate this we now present, in Figure 4, dielectric-scaled values of $\lambda_{exp}$ for aqueous NaCl, ionic liquid solutions in propylene carbonate, and for pure ionic liquids (those where literature values of $\varepsilon_r$ are

available[22-23]). For aqueous NaCl solutions, $\varepsilon_r$ depends strongly on concentration in the range 80.1 (pure water) to ~42 (5M NaCl)[24-25]. For mixtures of ionic liquid with propylene carbonate, $\varepsilon_r$ was calculated using Effective Medium Theory[26]. It is apparent in Figure 4 that the measured decay lengths in all electrolytes studied appears to collapse onto a single curve, providing a unique relationship between the deviation from the Debye length, $\lambda_{exp}/\lambda_D$, and the dielectric-scaled ion density as captured by the parameter $d/\lambda_D$ (where $d$ is the mean ion diameter in the electrolyte and so is constant for each electrolyte). The quantity $d/\lambda_D$ can be interpreted as a dielectric-scaled concentration because $1/\lambda_D$ is proportional to $(c^{1/2})/(\varepsilon^{1/2})$. The dielectric constant depends on the background solvent and the polarisability of the ions, and thus could be different even for ionic solutions with the same ion concentration and ion diameter. Therefore, for any particular salt and solvent combination $d/\lambda_D$ depends mainly on concentration, but in a way which is modulated by the dielectric constant which also varies as the ion concentration increases. This useful dimensionless quantity allows comparison of measured lengthscales in electrolytes with very different dielectric constants. Additionally, there is an implicit prediction of how temperature should affect the results since it is accounted for in $\lambda_D$. Mean ion diameter, $d$, is estimated from the cube root of the volume per ion pair halved, which gives values in good agreement with approximately half of the measured spacing in the oscillatory region of the force profiles for ionic liquids. Our method agrees well if the force oscillations arise from combined cation and anion layers, as is currently accepted to be the structure in ionic liquids at charged surfaces. A similar value of $d$ can also be inferred from the cation-anion correlation peak in x-ray scattering experiments with the same ionic liquid[27]. This is likely to be the most accurate, however there is limited number of ionic liquids studied in the literature. We note that alternative methods could be used to estimate or calculate the mean ion diameter, $d$, for example taking into account the packing fraction of ions, and it is not immediately clear whether the 'bare' or hydrated diameter should be taken for aqueous NaCl electrolytes. It will be important in future work to compare in detail the impact of different methods for assignment of $d$ for both ionic liquids and simple electrolytes in a systematic way. All quantities used in preparing Figure 4 are provided in the supporting information.

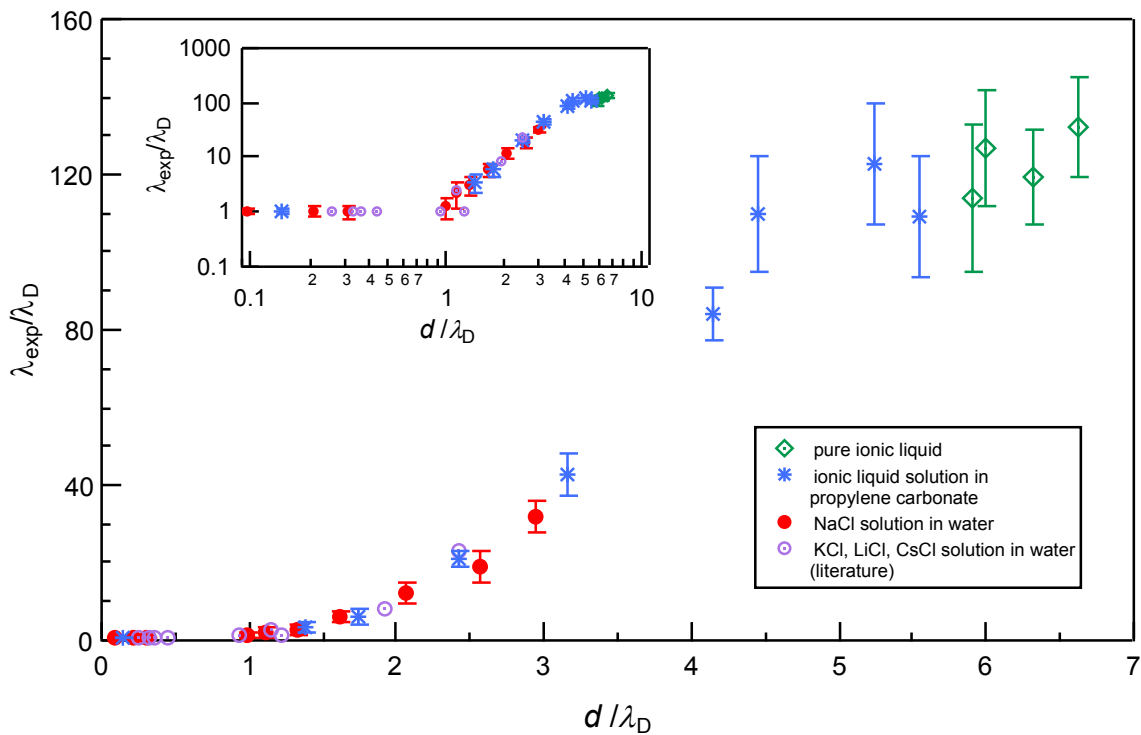

**Figure 4**
Deviation of the experimental decay length from the Debye length, $\lambda_{exp}/\lambda_D$, plotted vs. $d/\lambda_D$ for pure ionic liquids (open diamonds), aqueous NaCl solutions (filled circles), solutions of an ionic liquid, [$C_4C_1$Pyrr][$NTf_2$], mixed with propylene carbonate (asterisk symbols). Literature values[17, 19] are also included for aqueous LiCl, KCl, and CsCl solutions (open circles), the details of which are in the supporting information. Data is plotted for ionic liquids for which values of $\varepsilon_r$ are in the literature, and are in the range 12-16. Values of $\varepsilon_r$ for ionic liquid solutions in propylene carbonate were calculated using Effective Medium Theory[26], while for aqueous NaCl solutions $\varepsilon_r$ varies between 80.1 (pure water) and ~42 (5M NaCl)[24]. The inset shows the same data on a double logarithmic plot. The double logarithmic plot demonstrates the apparent power-law dependence in the range $1 < d/\lambda_D < 4$. All quantitative data and molecular structures relating to this figure are detailed in the supporting information.

The fact that the screening length of ionic liquids and simple electrolytes appears to collapse onto a single curve (Figure 4) suggests a striking degree of similarity between ionic liquids and concentrated salt solutions; it appears that microphase separation and bulk nanostructuring that occurs in many ionic liquids due to interactions between non-polar domains[28-33], or effects of ion size and shape anisotropy, do not alter the qualitative behaviour of the interionic interactions.

Although we do not intend to propose a theoretical model for the molecular origin of our observed trend in asymptotic decay with concentration here, we note that our observations are likely to be related to the works of Evans[8] and Attard[10] in the 1990's, where the mean spherical approximation (MSA) and hypernetted chain approximation (HNC) were used to study the transitions from monotonic to charge-oscillatory to density-oscillatory regimes with increasing electrolyte concentration. In each of these works a qualitatively very similar non-monotonic trend of $\lambda_{exp}$ with $d/\lambda_D$ was predicted, with the minimum in decay length occurring at the Kirkwood line; the point at which the charge-charge correlation function transitions from monotonic to oscillatory. The Kirkwood line is predicted to be at $d/\lambda_D \approx \sqrt{2}$; close but not identical to the point where the electrostatic screening length begins to increase with concentration in Figure 4. However we note substantial quantitative difference between the magnitude of the screening lengths measured here and the predictions of MSA and HNC.

Our experiments are carried out using two atomically smooth and macroscopic mica sheets as the substrates for force measurement across electrolytes, and it is worth commenting upon the generality or otherwise of forces measured between mica sheets. Mica is negatively charged in electrolyte solutions[34], causing a perturbation to liquid structure in the near-surface region. This, in turn, causes variations in pressure in the thin film between the surfaces relative to the bulk liquid, as detected in our force measurement. The magnitude of the force, and details of the oscillatory short-range part of the force, are therefore likely to be altered in their details when the substrate is changed. However the *screening length* of the monotonic part of the interaction, $\lambda_{exp}$ is expected to be a *function of the bulk liquid* only, as is the case for the Debye length, and thus insensitive to the nature of the surface.

The present experiments encourage comparison with several recent theoretical and experimental work. First, it is likely that the recently reported voltage-induced phase transition in ionic liquid-metal interfacial capacitance[35] and maximum in differential capacitance as a function of ionic liquid concentration[36] could be a part of a more general phenomenon also observable in concentrated electrolytes. Although this phase transition and anomaly in capacitance can be captured by positing a non-convex bulk free energy[37] or a charge-frustrated Ising model[36] which model interactions specific to ionic liquids, it remains to be seen whether there is a more general theoretical description for this phenomenon. Secondly, the demonstration of general non-monotonic dependence of decay length on electrolyte concentration should contribute towards interpreting the long-range forces detected in ionic liquids[11]; a theory for long range forces in ionic liquids must be consistent with our observed increase of screening length with concentration over a certain range. On the other hand, theories relating to the interplay between bulk and surface nanostructure[38] need to keep in mind the observation that similar long-range forces occur in NaCl solutions where nanostructuring due to solvophobic interactions is not expected to occur. However bulk oscillatory structuring is nonetheless expected in simple electrolytes, such as NaCl, at sufficiently high concentrations[39-40]

Our direct measurements of the force between macroscopic and atomically smooth mica sheets across electrolytes have revealed a clear non-monotonic dependence of decay length on concentration of the electrolyte. Within the Debye-Hückel region, below ca. 0.1 M, the screening length decreases with concentration according to the Debye theory reaching a minimum when the ion diameter matches the Debye length; $d/\lambda_D \sim 1$. At higher concentrations, in the range $1 < d/\lambda_D < 5$, the experimental decay length increases strongly with increasing concentration up to $(\lambda_{exp}/\lambda_D) \sim 120$. Deviation of the experimental decay length from the theoretical Debye length, $\lambda_{exp} / \lambda_D$, is described by the single dimensionless parameter, $d/\lambda_D$, for all electrolytes and ionic liquids studied, indicating a common mechanism for 1:1 electrolytes. Such substantial deviation from the canonical understanding of electrolytes is likely to be important for understanding electric fields and surface forces in concentrated electrolytes in a multitude of scenarios. These include intermolecular forces between interacting proteins and membranes in confined geometries where ion concentration can be high; and in energy storage applications, such as batteries and supercapacitors, where highly concentrated electrolytes are confined within porous electrodes.

**ACKNOWLEDGEMENTS**

This work was supported by The Leverhulme Trust (RPG-2015-328), the ERC (under Starting Grant LIQUISWITCH), and The John Fell Fund (Oxford University). A.M.S. was supported by a Doctoral Prize from the EPSRC. A.A.L. was supported by a Fulbright Fellowship to Harvard University. We are grateful to Lorna Swift and Tom Welton for synthesis of a phosphonium ionic liquid. We thank Bob Evans for illuminating discussions. A.A.L. appreciates fruitful discussions with David Limmer.

**Supporting Information**

Details of the materials and methods; details of the ionic liquids used and the data points used to plot Figure 3; quantities used to performing scaling in Figure 4 and the final data points in numerical form used to plot Figure 4 and including the literature values; discussion of alternative calculations of *d* and figures demonstrating how this impacts on $d/\lambda_D$.

# REFERENCES

(1.) Evans, D. F.; Wennerström, H. *The Colloidal Domain: Where Physics, Chemistry, Biology, and Technology Meet*; Wiley, 1999.

(2) Levin, Y. Electrostatic Correlations: From Plasma to Biology. *Reports on Progress in Physics* **2002**, *65*, 1577.

(3) Jing, Y.; Jadhao, V.; Zwanikken, J. W.; Olvera de la Cruz, M. Ionic Structure in Liquids Confined by Dielectric Interfaces. *J. Chem. Phys* **2015**, *143*, 194508.

(4) Yochelis, A.; Singh, M. B.; Visoly-Fisher, I. Coupling Bulk and near-Electrode Interfacial Nanostructuring in Ionic Liquids. *Chem. Mat.* **2015**, *27*, 4169-4179.

(5) Bockris, J. O. M.; Reddy, A. K. N. Volume 1: Modern Electrochemistry; Springer US, **1998**.

(6) Onsager, L., Theories of Concentrated Electrolytes. *Chem. Rev.* **1933**, *13*, 73-89.

(7) Kjellander, R.; Mitchell, D. J. An Exact but Linear and Poisson—Boltzmann-Like Theory for Electrolytes and Colloid Dispersions in the Primitive Model. *Chem. Phys. Lett.* **1992**, *200*, 76-82.

(8) Leote de Carvalho, R. J. F.; Evans, R. The Decay of Correlations in Ionic Fluids. *Mol. Phys.* **1994**, *83*, 619-654.

(9) Stillinger, F. H.; Lovett, R. Ion-Pair Theory of Concentrated Electrolytes. I. Basic Concepts. *J. Chem. Phys* **1968**, *48*, 3858-3868.

(10) Attard, P. Asymptotic Analysis of Primitive Model Electrolytes and the Electrical Double Layer. *Physical Review E* **1993**, *48*, 3604-3621.

(11) Gebbie, M. A.; Dobbs, H. A.; Valtiner, M.; Israelachvili, J. N. Long-Range Electrostatic Screening in Ionic Liquids. *Proc. Natl. Acad. Sci. USA* **2015**, *112*, 7432-7437.

(12) Smith, A. M.; Lovelock, K. R. J.; Perkin, S. Monolayer and Bilayer Structures in Ionic Liquids and Their Mixtures Confined to Nano-Films. *Farad. Discuss.* **2013**, *167*, 279-292.

(13) Pashley, R. M. Dlvo and Hydration Forces between Mica Surfaces in Li+,Na+,K+, and Cs+ Electrolyte Solutions: - a Correlation of Double-Layer and Hydration Forces with Surface Cation-Exchange Properties. *J. Colloid Interface Sci.* **1981**, *83*, 531-546.

(14) Perkin, S.; Crowhurst, L.; Niedermeyer, H.; Welton, T.; Smith, A. M.; Gosvami, N. N. Self-Assembly in the Electrical Double Layer of Ionic Liquids. *Chem. Commun.* **2011**, *47*, 6572-6574.

(15) Smith, A. M.; Lovelock, K. R. J.; Gosvami, N. N.; Licence, P.; Dolan, A.; Welton, T.; Perkin, S. Monolayer to Bilayer Structural Transition in Confined Pyrrolidinium-Based Ionic Liquids. *J. Phys. Chem. Lett.* **2013**, *4*, 378-382.

(16) Espinosa-Marzal, R. M.; Drobek, T.; Balmer, T.; Heuberger, M. P. Hydrated-Ion Ordering in Electrical Double Layers. *Phys. Chem. Chem. Phys.* **2012**, *14*, 6085-6093.

(17) Pashley, R. M.; Israelachvili, J. N. Molecular Layering of Water in Thin Films between Mica Surfaces and Its Relation to Hydration Forces. *J. Colloid Interface Sci.* **1984**, *101*, 511-523.

(18) Pashley, R. M. Forces between Mica Surfaces in La3+ and Cr3+ Electrolyte Solutions. *J. Colloid Interface Sci.* **1984**, *102*, 23-35.

(19) Baimpos, T.; Shrestha, B. R.; Raman, S.; Valtiner, M. Effect of Interfacial Ion Structuring on Range and Magnitude of Electric Double Layer, Hydration, and Adhesive Interactions between Mica Surfaces in 0.05–3 M Li+ and Cs+ Electrolyte Solutions. *Langmuir* **2014**, *30*, 4322-4332.

(20) Espinosa-Marzal, R. M.; Arcifa, A.; Rossi, A.; Spencer, N. D. Microslips to "Avalanches" in Confined, Molecular Layers of Ionic Liquids. *J. Phys. Chem. Lett.* **2013**, *5*, 179-184.

(21) Espinosa-Marzal, R. M.; Arcifa, A.; Rossi, A.; Spencer, N. D. Ionic Liquids Confined in Hydrophilic Nanocontacts: Structure and Lubricity in the Presence of Water. *J. Phys. Chem. C* **2014**, *118*, 6491-6503.

(22) Wakai, C.; Oleinikova, A.; Ott, M.; Weingärtner, H. How Polar Are Ionic Liquids? Determination of the Static Dielectric Constant of an Imidazolium-Based Ionic Liquid by Microwave Dielectric Spectroscopy. *J. Phys. Chem. B* **2005**, *109*, 17028-17030.

(23) Huang, M.-M.; Jiang, Y.; Sasisanker, P.; Driver, G. W.; Weingärtner, H. Static Relative Dielectric Permittivities of Ionic Liquids at 25 °C. *Journal of Chemical & Engineering Data* **2011**, *56*, 1494-1499.

(24) Buchner, R.; Hefter, G. T.; May, P. M. Dielectric Relaxation of Aqueous NaCl Solutions. *The Journal of Physical Chemistry A* **1999**, *103*, 1-9.

(25) Gavish, N.; Promislow, K. Dependence of the Dielectric Constant of Electrolyte Solutions on Ionic Concentration. *arXiv preprint arXiv:1208.5169* **2012**.

(26) Bergman, D. J. The Dielectric Constant of a Composite Material—a Problem in Classical Physics. *Physics Reports* **1978**, *43*, 377-407.

(27) Santos, C. S.; Murthy, N. S.; Baker, G. A.; Castner, E. W. Communication: X-Ray Scattering from Ionic Liquids with Pyrrolidinium Cations. *J. Chem. Phys.* **2011**, *134*, 121101.

(28) Urahata, S. M.; Ribeiro, M. C. C. Structure of Ionic Liquids of 1-Alkyl-3-Methylimidazolium Cations: A Systematic Computer Simulation Study. *J. Chem. Phys* **2004**, *120*, 1855-1863.

(29) Wang, Y. T.; Voth, G. A. Unique Spatial Heterogeneity in Ionic Liquids. *J. Am. Chem. Soc* **2005**, *127*, 12192-12193.

(30) Canongia Lopes, J. N. A.; Pádua, A. A. H. Nanostructural Organization in Ionic Liquids. *J. Phys. Chem. B* **2006**, *110*, 3330-3335.

(31) Triolo, A.; Russina, O.; Bleif, H.-J.; Di Cola, E. Nanoscale Segregation in Room Temperature Ionic Liquids. *J. Phys. Chem. B* **2007**, *111*, 4641-4644.

(32) Russina, O.; Triolo, A.; Gontrani, L.; Caminiti, R.; Xiao, D.; Hines Jr., L. G.; Bartsch, R. A.; Quitevis, E. L.; Plechkova, N.; Seddon, K. R. Morphology and Intermolecular Dynamics of 1-Alkyl-3-Methylimidazolium Bis{(Trifluoromethane)Sulfonyl}Amide Ionic Liquids: Structural and Dynamic Evidence of Nanoscale Segregation. *Journal of Physics: Condensed Matter* **2009**, *21*, 424121.

(33) Shimizu, K.; Costa Gomes, M. F.; Pádua, A. A.; Rebelo, L. P.; Canongia Lopes, J. N. Three Commentaries on the Nano-Segregated Structure of Ionic Liquids. *Journal of Molecular Structure: THEOCHEM* **2010**, *946*, 70-76.

(34) Perkin, S.; Albrecht, T.; Klein, J. Layering and Shear Properties of an Ionic Liquid, 1-Ethyl-3-Methylimidazolium Ethylsulfate, Confined to Nano-Films between Mica Surfaces. *Phys. Chem. Chem. Phys.* **2010**, *12*, 1243-1247.

(35) Merlet, C.; Limmer, D. T.; Salanne, M.; van Roij, R.; Madden, P. A.; Chandler, D.; Rotenberg, B. The Electric Double Layer Has a Life of Its Own. *J. Phys. Chem. C* **2014**, *118*, 18291-18298.

(36) Bozym, D. J.; Uralcan, B.; Limmer, D. T.; Pope, M. A.; Szamreta, N. J.; Debenedetti, P. G.; Aksay, I. A. Anomalous Capacitance Maximum of the Glassy Carbon–Ionic Liquid Interface through Dilution with Organic Solvents. *J. Phys. Chem. Lett.* **2015**, *6*, 2644-2648.

(37) Limmer, D. T. Interfacial Ordering and Accompanying Divergent Capacitance at Ionic Liquid-Metal Interfaces. *Phys. Rev. Lett.* **2015**, *115*, 256102.

(38) Li, H.; Endres, F.; Atkin, R. Effect of Alkyl Chain Length and Anion Species on the Interfacial Nanostructure of Ionic Liquids at the Au(111)-Ionic Liquid Interface as a Function of Potential. *Phys. Chem. Chem. Phys.* **2013**, *15*, 14624-14633.

(39) Kirkwood, J. G. Statistical Mechanics of Liquid Solutions. *Chem. Rev.* **1936**, *19*, 275-307.

(40) Gavish, N.; Yochelis, A. Theory of Phase Separation and Polarization for Pure Ionic Liquids. *J. Phys. Chem. Lett.* **2016**, *7*, 1121-1126.

# Supporting Information

# The Electrostatic Screening Length in Concentrated Electrolytes Increases with Concentration

Alexander M. Smith*,[a], Alpha A. Lee*,[b] and Susan Perkin*,[a]

[a]Department of Chemistry, Physical & Theoretical Chemistry Laboratory, University of Oxford, Oxford OX1 3QZ, U.K.
[b]School of Engineering and Applied Sciences, Harvard University, Cambridge, MA 02138, USA

Corresponding author emails:

susan.perkin@chem.ox.ac.uk
alphalee@g.harvard.edu
alexander.smith@chem.ox.ac.uk

**Materials**

Mica was highest grade of the ruby muscovite variety (S&J Trading). The ionic liquids investigated here were supplied by: Iolitec ($[C_4C_1Pyrr][NTf_2]$, $[C_2C_1Im][NTf_2]$), Merck ($[C_2C_1Im][FAP]$), Sigma-Aldrich ($[C_2C_1Im][BF_4]$, $[C_2C_1Im][OTf]$, $[P_{6,6,6,14}][NTf_2]$), or synthesised via modified procedures from established synthetic methods[1] ($[P_{2,2,2,8}][NTf_2]$). The structures of these ionic liquids are shown in Table S1. All ionic liquids were dried in vacuo ($10^{-2}$ mbar, 70 °C) overnight before use. Sodium chloride (Sigma-Aldrich, 99.999%) and ultrapure water (18.2 MΩ cm resistivity, <3 ppb TOC) were used for preparing aqueous electrolyte solutions. Propylene carbonate (Sigma Aldrich, anhydrous, 99.7%) was used from freshly opened bottles and mixed with $[C_4C_1Pyrr][NTf_2]$. For experiments with non-aqueous electrolytes, additional care was taken to keep the atmosphere inside the apparatus chamber dry by purging with dry nitrogen, and a vial containing $P_2O_5$ was placed near the surfaces to absorb any residual water vapour. For reasons of solubility, EPON 1004 (Shell Chemicals) was used to glue mica pieces for aqueous electrolytes and ionic liquid rich solutions, while glucose (Sigma-Aldrich, 99.5%) is used for the propylene carbonate rich solutions. For the intermediate propylene carbonate concentrations, care was taken to use as large mica pieces as possible and small drops of liquid injected to avoid contact between the liquid and glue. Contamination due to dissolved glue diffusing within the contact region is readily apparent as enhanced viscosity and strong repulsion measured at large distances, obscuring any molecular layering.

**Methods**

Detailed experimental procedures for surface force measurements are described in detail elsewhere[2]. Two back-silvered (~40 nm) mica pieces of uniform thickness (1-3 µm in these experiments) are glued to cylindrical silica lenses with the silver side down. The lenses are mounted in the apparatus facing each other in a crossed-cylinder configuration, where the mica surfaces have a point of closest approach geometrically equivalent to a sphere near a flat surface. White-light multiple beam interferometry is used to determine $D$ by means of constructive interference fringes of equal chromatic order (FECO)[3] which are observed using a spectrometer and captured with a CCD camera. The deflection of a horizontal leaf spring is measured directly via interferometry to determine the normal force, $F_N$, with resolution better than $10^{-7}$ N. Forces are normalised by the mean local radius of curvature, $R$, of the glued mica surfaces ($R \approx 1$ cm) inferred from the interference pattern and allows comparison between different contact regions. It also provides a way to compare with theoretical values for the interaction energy, $E$, between flat surfaces at the same surface separation using the Derjaguin approximation, $F_N/R = 2\pi E$. Surfaces were approached at constant velocity (0.5 – 3 nm s$^{-1}$) from $D$ ~300 nm (where $F_N = 0$) down to molecularly confined films using a piezoelectric tube upon which one of the surfaces is mounted, and a CCD camera was used to capture the FECO at high frame rate. We note that such dynamic approaches improve the force resolution compared to experiments where each data point is recorded manually using the spectrometer eyepiece[4] and thermal drift is more problematic, particularly for the more viscous liquids.

**Table S1 Experimental decay lengths for pure ionic liquids**

| Ionic liquid | Structure | Abbreviation | Concentration (M) | $\lambda_{exp}$ (nm) |
|---|---|---|---|---|
| 1-butyl-1-methylpyrrolidinium bis[(trifluoromethane)sulfonyl]imide | | [C$_4$C$_1$Pyrr][NTf$_2$] | 3.31 | 8.4 ± 0.9 |
| 1-ethyl-3-methylimidazolium bis[(trifluoromethane)sulfonyl]imide | | [C$_2$C$_1$Im][NTf$_2$] | 3.91 | 7.1 ± 0.7 |
| 1-ethyl-3-methylimidazolium tris[(pentafluoroethane)trifluorophosphate | | [C$_2$C$_1$Im][FAP] | 3.07 | 9.2 ± 0.6 |
| 1-ethyl-3-methylimidazolium tetrafluoroborate | | [C$_2$C$_1$Im][BF$_4$] | 6.54 | 6.3 ± 0.6 |
| 1-ethyl-3-methylimidazolium trifluoromethanesulfonate | | [C$_2$C$_1$Im][OTf] | 5.34 | 6.6 ± 1.1 |
| triethyloctylphosphonium bis[(trifluoromethane)sulfonyl]imide | | [P$_{2,2,2,8}$][NTf$_2$] | 2.46 | 8.2 ± 0.9 |
| trihexyltetradecylphosphonium bis[(trifluoromethane)sulfonyl]imide | | [P$_{6,6,6,14}$][NTf$_2$] | 1.40 | 6.5 ± 1.4 |

**Table S2** Experimental decay lengths for solutions of NaCl in water

| Concentration (M) | $\lambda_{exp}$ (nm) |
|---|---|
| 0.01 | 3.0 ± 0.3 |
| 0.05 | 1.4 ± 0.3 |
| 0.10 | 0.9 ± 0.3 |
| 0.90 | 0.4 ± 0.2 |
| 1.12 | 0.6 ± 0.3 |
| 1.48 | 0.7 ± 0.2 |
| 2.04 | 1.1 ± 0.3 |
| 2.98 | 1.7 ± 0.4 |
| 4.06 | 2.2 ± 0.5 |
| 4.93 | 3.2 ± 0.4 |

**Table S3** Experimental decay lengths for solutions of [C$_4$C$_1$Pyrr][NTf$_2$] in propylene carbonate

| Concentration (M) | $\lambda_{exp}$ (nm) |
|---|---|
| 0.01 | 2.7 ± 0.3 |
| 0.70 | 1.0 ± 0.4 |
| 0.98 | 1.5 ± 0.4 |
| 1.50 | 3.4 ± 0.4 |
| 1.97 | 5.4 ± 0.7 |
| 2.50 | 8.1 ± 0.7 |
| 2.65 | 9.8 ± 1.3 |
| 3.00 | 9.3 ± 1.2 |
| 3.13 | 7.8 ± 1.1 |

**Table S4** Values of $d$ and $\varepsilon_r$ for ionic liquids used to prepare Figure 4

| Ionic liquid | $d$ (nm) | $\varepsilon_r$ |
|---|---|---|
| [C$_4$C$_1$Pyrr][NTf$_2$] | 0.40 | 12.5 |
| [C$_2$C$_1$Im][NTf$_2$] | 0.38 | 12 |
| [C$_2$C$_1$Im][BF$_4$] | 0.32 | 12.9 |
| [C$_2$C$_1$Im][OTf] | 0.34 | 15.2 |

**Table S4** Values of $d$ and $\varepsilon_r$ for solutions of NaCl in water used to prepare Figure 4

| Concentration (M) | $d$ (nm) | $\varepsilon_r$ |
|---|---|---|
| 0.01 | 0.294 | 80 |
| 0.05 | 0.294 | 79 |
| 0.10 | 0.294 | 78 |
| 1.12 | 0.294 | 68 |
| 1.48 | 0.294 | 63 |
| 2.04 | 0.294 | 58 |
| 2.98 | 0.294 | 52 |
| 4.06 | 0.294 | 46 |
| 4.93 | 0.294 | 43 |

**Table S5** Values of $d$ and $\varepsilon_r$ for solutions of [C$_4$C$_1$Pyrr][NTf$_2$] in propylene carbonate used to prepare Figure 4

| Concentration (M) | $d$ (nm) | $\varepsilon_r$ |
|---|---|---|
| 0.01 | 0.40 | 64 |
| 0.98 | 0.40 | 44 |
| 1.50 | 0.40 | 35 |
| 1.97 | 0.40 | 27 |
| 2.50 | 0.40 | 20 |
| 2.65 | 0.40 | 18 |
| 3.00 | 0.40 | 15 |
| 3.13 | 0.40 | 14 |

**Table S6** Values of $d$ and $\varepsilon_r$ for KCl, LiCL and CsCl solutions used to prepare Figure 4

| Solution | $d$ (nm) | $\varepsilon_r$ |
|---|---|---|
| 1 M KCl | 0.318 | 67 |
| 0.15 M LiCl | 0.257 | 78 |
| 1 M LiCl | 0.257 | 65 |
| 3 M LiCl | 0.257 | 46 |
| 0.05 M CsCl | 0.348 | 79 |
| 0.1 M CsCl | 0.348 | 79 |
| 0.15 M CsCl | 0.348 | 78 |
| 1 M CsCl | 0.348 | 70 |
| 3 M CsCl | 0.348 | 53 |

In Figure 4 of our letter, values for the ion diameter, $d$, were estimated from the molar mass and density of ionic liquids, and correspond to the average *unhydrated* ion diamater for NaCl. However there are different ways to calculate $d$. For example, $d$ for ionic liquids may instead be inferred from the cation-anion correlation peak in x-ray scattering experiments[5]. This is shown in Figure S1. The data still appears to collapse onto a single curve, although the specific values of $d/\lambda_D$ differ slightly. Figure S2 shows the effect of instead using the *hydrated* ion diameter for NaCl. The data no longer fits on a single curve, although it is not clear how a fully hydrated ion can apply for such high concentrations in confined films.

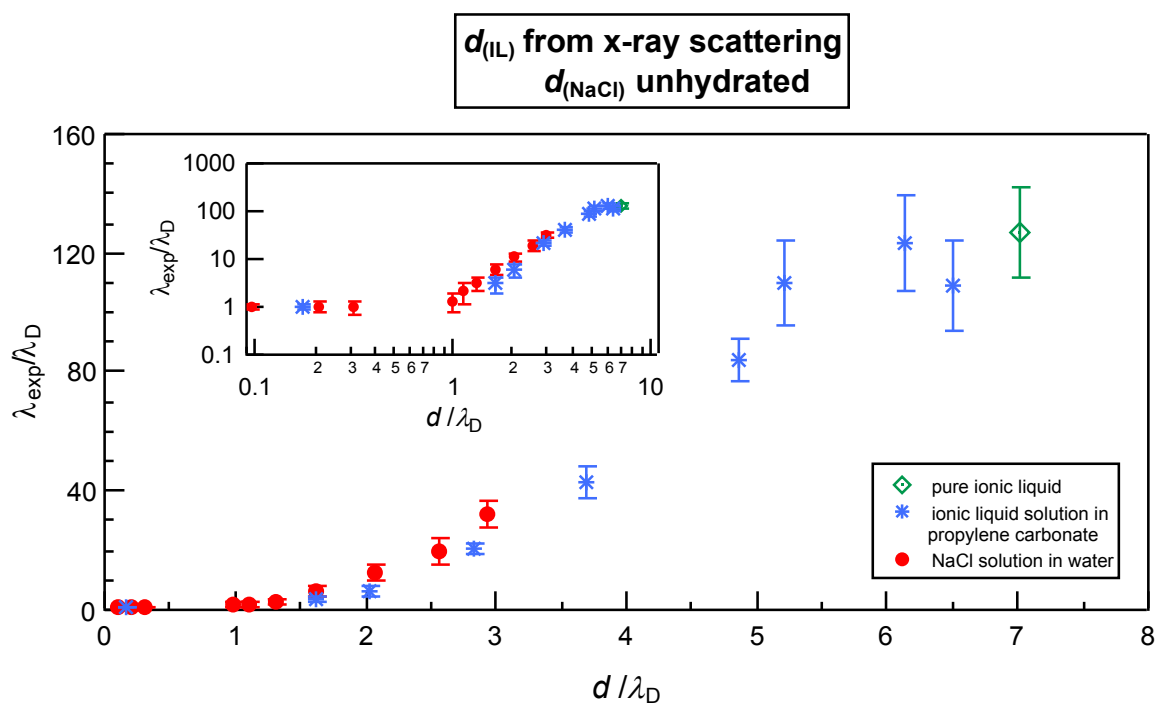

**Figure S1**
Deviation of the experimental decay length from the Debye length, $\lambda_{exp}/\lambda_D$ plotted vs. $d/\lambda_D$ for the pure ionic liquid [$C_4C_1$Pyrr][$NTf_2$] (open symbol), aqueous NaCl solutions (closed symbols), and the ionic liquid [$C_4C_1$Pyrr][$NTf_2$] mixed with propylene carbonate (asterisk symbols). Values for $d$ were inferred from x-ray scattering experiments[5] for the ionic liquid (0.465 nm) and correspond to the average *unhydrated* ion diameter for NaCl (0.294 nm). The inset shows the same data on a double logarithmic plot.

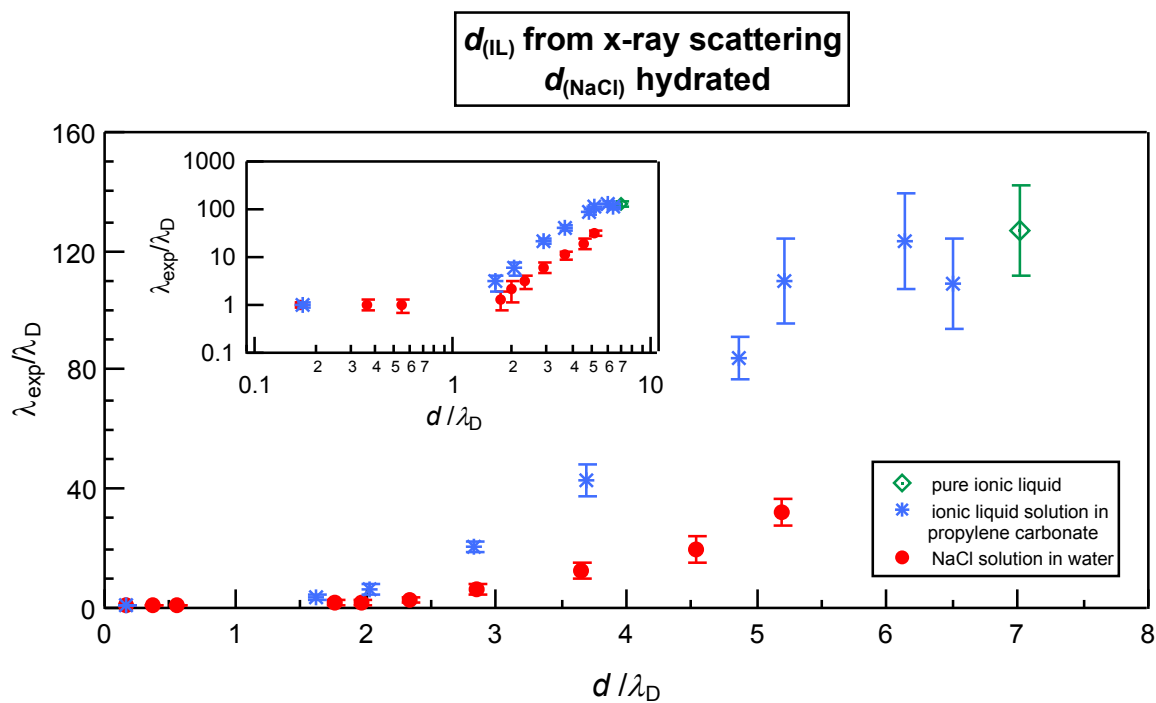

**Figure S2**

Deviation of the experimental decay length from the Debye length, $\lambda_{exp}/\lambda_D$ plotted vs. $d/\lambda_D$ for the pure ionic liquid $[C_4C_1Pyrr][NTf_2]$ (open symbol), aqueous NaCl solutions (closed symbols), and the ionic liquid $[C_4C_1Pyrr][NTf_2]$ mixed with propylene carbonate (asterisk symbols). Values for $d$ were inferred from x-ray scattering experiments[5] for the ionic liquid (0.465 nm) and correspond to the average *hydrated* ion diameter for NaCl (0.52 nm). The inset shows the same data on a double logarithmic plot.

## References

(1) Ab Rani, M. A.; Brant, A.; Crowhurst, L.; Dolan, A.; Lui, M.; Hassan, N. H.; Hallett, J. P.; Hunt, P. A.; Niedermeyer, H.; Perez-Arlandis, J. M.; Schrems, M.; Welton, T.; Wilding, R. Understanding the Polarity of Ionic Liquids. *Phys. Chem. Chem. Phys.* **2011**, *13*, 16831-16840.
(2) Klein, J.; Kumacheva, E. Simple Liquids Confined to Molecularly Thin Layers. I. Confinement-Induced Liquid-to-Solid Phase Transitions. *J. Chem. Phys.* **1998**, *108*, 6996-7009.
(3) Israelachvili, J. Thin-Film Studies Using Multiple-Beam Interferometry. *J. Colloid Interface Sci.* **1973**, *44*, 259-272.
(4) Perkin, S. Ionic Liquids in Confined Geometries. *Phys. Chem. Chem. Phys.* **2012**, *14*, 5052-5062.
(5) Santos, C. S.; Murthy, N. S.; Baker, G. A.; Castner, E. W. Communication: X-Ray Scattering from Ionic Liquids with Pyrrolidinium Cations. *J. Chem. Phys.* **2011**, *134*, 121101.